\documentclass[aps,prl,reprint,groupedaddress,showpacs]{revtex4-1}
\usepackage{graphicx}
\usepackage{amssymb}
\usepackage{epstopdf}
\usepackage{amsmath}
\usepackage{pifont}

\usepackage{hyperref} %uncomment these two lines when using pdflatex to get links in article. NB only latex likes this package not tex!
\begin{document}

\title{Excitation of weakly-bound molecules to trilobite-like Rydberg states.}

\author{M. A. Bellos}
\altaffiliation{Present Address: Yale University, Department of Physics, P.O. Box 208120, New Haven, Connecticut 06520-8120, USA}
\author{R. Carollo}
\author{J. Banerjee}
\author {E. E. Eyler}
\author{P. L. Gould}
\author{W. C. Stwalley}
\affiliation{Department of Physics, University of Connecticut, Storrs, Connecticut 06269-3046, USA}

\date{\today}

\begin{abstract}
We observe ``trilobite-like" states of ultracold $^{85}$Rb$_2$ molecules, in which a ground-state atom is bound by the electronic wavefunction of its Rydberg-atom partner. We populate these states through the ultraviolet excitation of weakly-bound molecules, and access a regime of trilobite-like states at low principal quantum numbers and with vibrational turning points around 35 Bohr radii. This demonstrates that, unlike previous studies that used free-to-bound transitions, trilobite-like states can also be excited through bound-to-bound transitions. This approach provides high excitation probabilities without requiring high-density samples, and affords the ability to control the excitation radius by selection of the initial-state vibrational level.\\
\end{abstract}


\maketitle
A class of long-range Rydberg molecules, sometimes called ``trilobite molecules," occurs when a ground-state atom is embedded within the electronic wavefunction of a Rydberg atom \cite{greene00}. The bond between the Rydberg atom and the ground-state atom originates from the attractive interaction between the Rydberg electron and the ground-state atom \cite{greene00}. This bond has been described as a new type of chemical bond, distinct from the well-known covalent, ionic, and van der Waals bonds \cite{hogan09}. The name ``trilobite molecule" was coined because in certain states, the perturbed Rydberg-electron wavefunction resembles a trilobite fossil \cite{greene00}. Trilobite states are characterized by large, and degenerate, values of orbital angular momentum ($l\geq3$) and large permanent electric dipole moments (EDMs $\sim 1\,$kDebye). Although pure trilobite states have yet to be observed, \textit{trilobite-like} states, bound by the same novel chemical bond but characterized primarily by lower values of $l$ and smaller EDMs, have been observed at ultracold temperature in photoassociation to bound vibrational levels \cite{bendkowsky09,bendkowsky10,butscher10,butscher11,li11,tallant12}, and at high temperatures in the form of satellite structures in the wings of atomic transitions \cite{greene06,vadla09}.

We adopt the name ``trilobite-like" molecules for the low-$l$ states to distinguish them from other types of long-range Rydberg molecules, such as macrodimers \cite{boisseau02,farooqi03,overstreet09} or heavy Rydberg atoms \cite{reinhold05,kirrander13}.

The bond length of a trilobite-like molecule can vary greatly depending on the principal quantum number, due to the $n^2$ dependence of the radius of the outermost lobe of the Rydberg wavefunction. For instance, the vibrational outer turning points of trilobite-like molecules have ranged from below 150 Bohr radii ($a_0$) in Refs. \cite{greene06,vadla09} to above 10$^3\,a_0$ in Refs. \cite{bendkowsky09,bendkowsky10,butscher10,butscher11,li11,tallant12}.

Here we report the observation of trilobite-like states of $^{85}$Rb$_2$ by a new method.  We start by producing ultracold Rb$_2$ molecules in a high vibrational level of the metastable $a\,^3\Sigma_u^+$ state via photoassociation of atoms in a magneto-optical trap (MOT), then excite them directly to trilobite-like states, detecting them via their autoionization into Rb$_2^+$ molecular ions. We observe these states in some of the lowest principal quantum numbers for which they can exist, $n\,=\,7$ and 9--12, and also for some of the shortest internuclear separations at which they can form.  As far as we are aware, this is the first time that trilobite-like states have been produced via bound-bound molecular transitions.

The range we have investigated overlaps with previous studies conducted in heat-pipe ovens \cite{greene06,vadla09}, in which satellite structures in the line wings of collisionally broadened Rb were found to correspond closely with minima of calculated long-range potential energy curves (PECs) \cite{greene06}. Although we access some of the same PECs, there are notable differences in the excitation pathway, detection scheme, and the attainable linewidth.

The low-$l$ trilobite-like binding energy (in atomic units) between the Rydberg-electron and the embedded atom can be approximated by

\begin{equation}\label{potentialenergy}
V(R)=2\pi \, a_s \, \vert \psi(R)  \vert^2  +6\pi \, a^3_p \, \vert \nabla \psi(R)  \vert^2,
\end{equation}
where $\psi(R)$ is the unperturbed Rydberg-electron wavefunction, and $a_s$ and $a_p$ are the energy-dependent (and therefore $R$-dependent) $s$- and $p$-wave scattering lengths, respectively, for low-energy collisions between the electron and the embedded atom \cite{omont77,hamilton02,bendkowsky10}. The first term of Eq.(\ref{potentialenergy}) originates from $s$-wave scattering, and the second from $p$-wave scattering.  In the low-$n$ small-$R$ regime of Rb$_2$, the second term is dominant as the $p$-wave scattering length is strongly influenced by the presence of a electron$\,$+$\,$atom shape resonance \cite{bahrim00,khuskivadze02,hamilton02}. However, for interactions in the high-$n$ large-$R$ regime the first term can dominate, such as when the kinetic energy of the Rydberg electron is too small to support $p$-wave collisions \cite{junginger12}.

\begin{figure}[]
\includegraphics[scale=0.69]{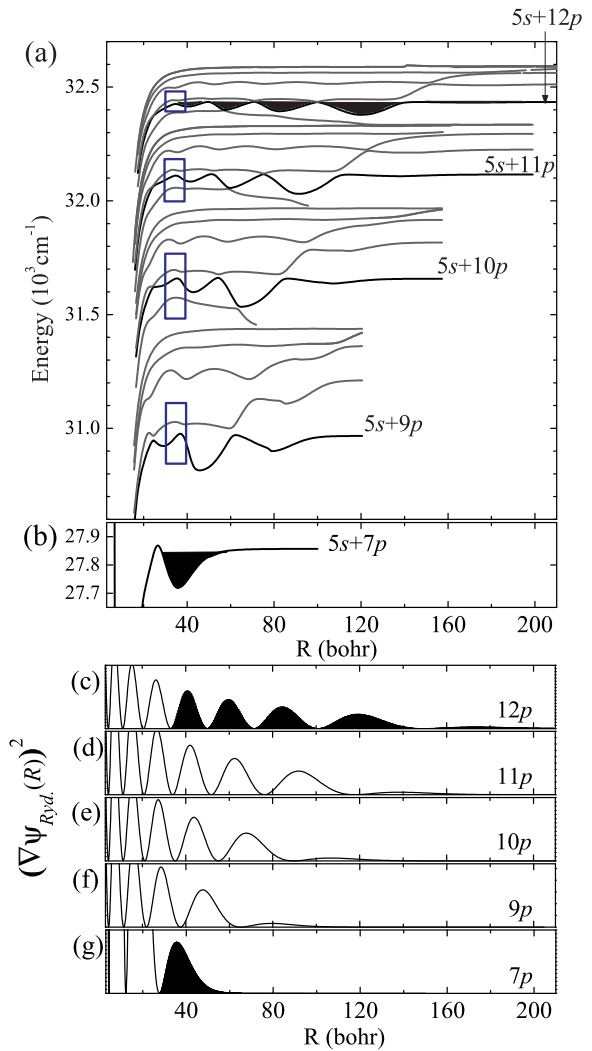}
\caption{\label{wavefucntions} (color online) (a) $^3\Sigma^+$ excited-state potential energy curves from Ref. \cite{greene06} calculated through Fermi pseudopotential methods. The rectangular boxes ($\square$) label the outer turning points of states populated in the present work, which are re-plotted in greater detail in Fig. \ref{spectra}. (b) $^3\Sigma_g^+$ excited-state potential energy curve from Ref. \cite{bellos13} calculated through \textit{ab initio} methods. (c)--(g) Derivatives of the radial Rydberg-electron wavefunctions for 12$p$, 11$p$, 10$p$, 9$p$ and 7$p$ atoms, respectively. The Rydberg-electron wavefunctions are approximated by phase-shifted hydrogenic wavefunctions calculated using Numerov integration in square-root coordinates \cite{bhatti81}. The shaded regions illustrate the correlation between potential wells and oscillations of the electron wavefunction derivatives.}
\end{figure}

For a Rydberg electron in quantum state $\left | l=1, m=0 \right>$, the $\vert \nabla \psi(R)  \vert^2$ term in Eq. (\ref{potentialenergy}) is proportional to $(\frac{\partial \psi(R)}{\partial R})^2$ and correlates to PECs of $^3\Sigma^+$ symmetry as shown in Fig. \ref{wavefucntions}. In the case of an electron in quantum state $\left | l=1, m=\pm1 \right>$, the $\vert \nabla \psi(R)  \vert^2$ term is instead proportional to $(\frac{\psi(R)}{R})^2$ and correlates to PECs of $^3\Pi$ symmetry. Such correlations between PECs and the Rydberg-electron wavefunctions also occur in the undulations of \textit{ab initio} excited-state PECs, showing that conventional \textit{ab initio} PECs can reproduce the low-$n$ trilobite-like bond \cite{yiannopoulou99}. In principle, the excited-state PECs of all molecules could exhibit these undulations, as long as the scattering length between the Rydberg electron and embedded atom is non-zero. In Rb$_2$, and other alkali-metal dimers such as Cs$_2$ and Fr$_2$, the existence of a $p$-wave triplet shape resonance \cite{bahrim00,khuskivadze02} produces large $p$-wave scattering lengths. As a result, Rb$_2$ has low-$n$ potential wells that can exceed 100$\,$cm$^{-1}$ in depth \cite{greene06}, and support over 100 vibrational levels. Unlike the case of high-$n$ large-$R$ trilobite-like states, a large number of vibrational levels are supported in each well and tunneling between wells is not significant due to the large barriers between wells.

In Fig. \ref{wavefucntions}(a) we plot the relevant excited-state PECs from Ref. \cite{greene06}, which were calculated using the Fermi pseudopotential method. These trilobite-like states, due to their EDMs, do not have \textit{ungerade}/\textit{gerade} symmetry \cite{greene00,li11}. The curves do not extend into the short-range region, because the Fermi pseudopotential model breaks down at separations where covalent bonding appears. Despite this breakdown, these Fermi pseudopotential curves are consistent with our observations, which are dominated by excitation around $R=35\,a_0$. We have converted the difference potentials of Ref. \cite{greene06} (which characterize free-bound excitation) back to full excited-state potentials (which characterize bound-bound excitation), by simply adding the long-range ground-state potential energy \footnote{For the ground-state potential, $V(R)\,=\,-\frac{C_6}{R^6}$, $C_6\,=\,4660$ a.u., see Ref. \cite{greene06}. The resulting correction between difference PECs and regular PECs is less than 0.5$\,$cm$^{-1}$ at $R=35\,a_0$.}. In Fig. \ref{wavefucntions}(b) we plot the relevant \textit{ab initio} excited-state PEC for the $5s+7p$ asymptote \cite{allouche12}. Observations of the low-$n$ medium-$R$ regime should allow detailed comparison of the \textit{ab initio} and Fermi pseudopotential methods at the extremes of their applicability.

The apparatus and procedure for initial state preparation have been previously described in Ref. \cite{bellos13} and are only briefly summarized here. The starting point is a MOT that traps about $8\times10^7$ $^{85}$Rb atoms at a peak density of $1\times10^{11}$ cm$^{-3}$ and a temperature of 120 $\mu$K. We form excited-state molecules by photoassociating atom pairs into the level $\left | 1(0_g^-), v'\simeq 173, J'=1\right>$ using a cw laser tuned to 796$\,$nm \cite{lozeille06}. We rely on spontaneous emission to produce weakly-bound $J=0$ and $J=2$ molecules, most of which populate the single vibrational level $\left | a\,^3\Sigma_u^+, v= 35 \right>$, with a binding energy of $E_B=-0.8\,\mathrm{cm}^{-1}$ relative to the $5s+5s$ ground-state atomic limit \cite{bellos13}. We use this weakly-bound level as the starting point for excitation to trilobite-like states. Although molecules are continuously produced in the MOT by the PA laser, they are also continuously lost because they are not trapped. As the molecules free-fall below the MOT, we photoexcite them with a pulsed uv laser and detect the formation of Rb$_2^+$ ions. A positively-charged electric field plate accelerates the Rb$_2^+$ ions from the center of the vacuum chamber towards an ion detector, where they are distinguished from Rb$^+$ ions by their time-of-flight. The charged field plate produces an electric field at the position of the MOT of about 100 V$\,$cm$^{-1}$. Pulsing the electric field on immediately after the photoexcitation does not lead to observable changes in signal strength, indicating that field ionization is not the dominant ionization mechanism.

We produce the uv light by frequency doubling an infrared pulsed dye laser, operated with a LDS750 laser dye, to access the energy region around the $5s+7p$ atomic limit, and with a DCM dye to access the higher atomic limits. We have not examined the energy region around $5s+8p$ because the required wavelengths are more difficult to produce. Our excitation laser is well-suited for broad survey scans but because its linewidth ($0.9\,$cm$^{-1}$) is comparable to typical excited-state vibrational spacings, we are only able to resolve individual vibrational levels in favorable cases. We anticipate that order-of-magnitude gains in resolution are possible through the use of a pulse-amplified cw laser \cite{salour77,eyler97}.

The production and detection of trilobite-like states can be summarized by the steps,
\begin{equation}\label{steps}
\mathrm{Rb}_2  + h\nu \xrightarrow{\mathrm{}} \mathrm{Rb}_2^{**} \xrightarrow{\mathrm{}} \mathrm{Rb}_2^+ + e^-.
\end{equation}
The first step is a single-photon bound-bound excitation, which corresponds to a change in bonding from an initial state bound primarily by van der Waals forces to a final state bound primarily by the trilobite-like bond of Eq. (\ref{potentialenergy}). The designation $\mathrm{Rb}_2^{**}$ refers to excited states of the molecule, either trilobite-like or non-trilobite-like, that are energetically above the ionization threshold and can spontaneously autoionize. We have previously deduced that our Rb$_2^+$ signals arise from autoionization rather than photoionization \cite{bellos13}. The fact that we observe any Rb$_2^+$ at all indicates that the autoionization lifetime is at least comparable to the radiative decay rate of the excited state.

Production of these trilobite-like states is quite efficient. With only a few hundred Rb$_2$ molecules present within the uv excitation beam, we detect up to 20 or 30 Rb$_2^+$ ions per pulse, implying a probability for excitation followed by autoionization (Eq. (\ref{steps})) of roughly 10\%. The high transition probability is partly due to the large vibrational wavefunction overlap between weakly-bound molecules and the long-range trilobite-like states. Because the initial state is a single bound molecule, there is no dependence on sample density.

\begin{figure}[]
\includegraphics[scale=0.8]{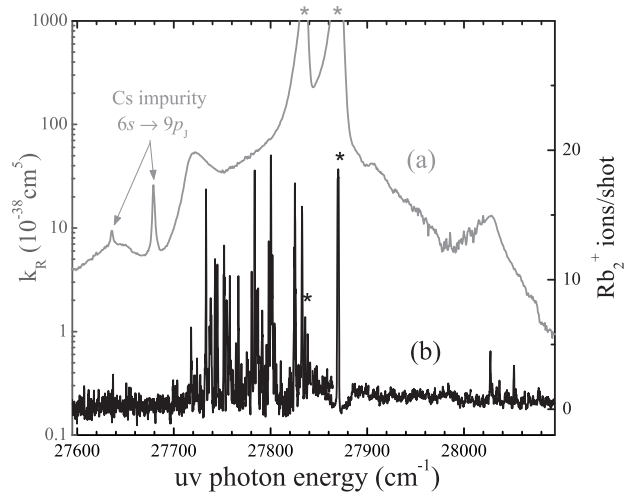}
\caption{\label{spectracomparison} (a) Reduced absorption coefficient of rubidium in a heat-pipe oven as a function of uv frequency, from Ref. \cite{greene06}. The lines labeled by two arrows are atomic transitions arising from cesium contamination and are not present in our experiment. (b) Rb$_2^+$ autoionization signal from laser excitation of high-$v$ ultracold molecules (present work). Both experiments access the same excited-state region around $5s+7p$. Lines marked with an asterisk ($\ast$) originate from the strong atomic transitions $5s\rightarrow7p_{\frac{1}{2}},7p_{\frac{3}{2}}$. In (b), these signals are due to leakage from the Rb$^+$ time-of-flight channel into the Rb$_2^+$ channel, and therefore significantly underestimate the actual Rb$^+$ ion production rate.}
\end{figure}

\begin{figure}[]
\includegraphics[scale=0.7]{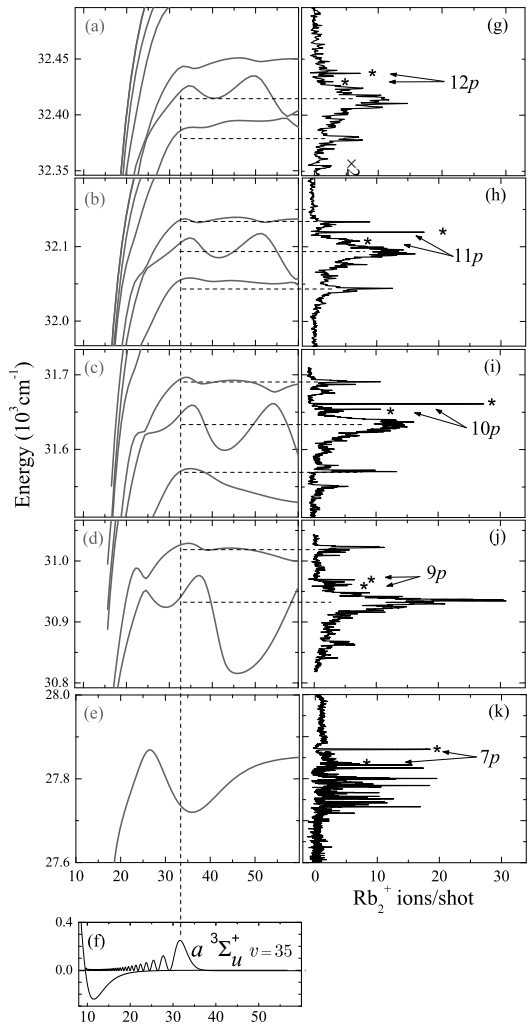}
\caption{\label{spectra} (a)--(e) Close-ups of the trilobite-like $^3\Sigma^+$ PECs around the $5s+np$, $n$= 12, 11, 10, 9 and 7 asymptotes, respectively, and corresponding Rb$_2^+$ ion spectra (g)--(k). (f) PEC and vibrational probability density of the initial $\left | a\,^3\Sigma_u^+, v= 35\right>$ state. The lines marked with an asterisk ($\ast$) denote atomic transitions from $5s$ to $np_{\frac{1}{2}},np_{\frac{3}{2}}$. The dashed vertical line depicts the center of the Franck-Condon window for the transition from the last lobe of the initial-state vibrational wavefunction to various trilobite-like PECs. The dashed horizontal lines indicate assignments of strong spectral features to the outer turning points of selected PECs. Some of the weak lines can be traced back to the second-to-last lobe of the initial-state vibrational wavefunction.}
\end{figure}

In Fig. \ref{spectracomparison}(b) we plot the number of Rb$_2^+$ ions detected per uv pulse as a function of the applied uv frequency.
Broad similarities are evident between the present work and previous heat-pipe spectra \cite{greene06,vadla09}, plotted in Fig. \ref{spectracomparison}(a). These similarities exist despite differences in the initial conditions (hot colliding atoms vs. ultracold molecules), detection mechanism (absorption vs. ionization), and temperature (1000$\,$K vs. 10$^{-4}\,$K).
In Figs. \ref{spectra}(a)--(e) we plot the inner-most wells of trilobite-like states and provide an assignment between PECs and observed spectral features in Figs. \ref{spectra}(g)--(k).
The broad similarities between the spectra in the present work and the structure seen previously in a heat-pipe oven \cite{greene06,vadla09} can still be observed for the lines blue of the atomic transitions to $9p$--$11p$, but not for lines red of the atomic transitions. At least some differences between our spectra and the heat-pipe observations is expected for all asymptotes above $5s+7p$, because in the present work we excite to trilobite-like wells with internuclear separations less than $R=40\,a_0$. In a heat-pipe oven, free-bound transitions to any trilobite-like well can occur. Despite the difference in Franck-Condon overlap between bound-bound transitions and free-bound transitions, we observe that lines to the blue of the atomic transitions to $9p$--$11p$ match the energies of features observed in a heat-pipe oven. Interestingly, in Ref. \cite{vadla09} the feature blue of 11$p$ is attributed to trilobite-like wells with $R\geq63\,a_0$; however, in the present work, the line blue of the 11$p$ asymptote (Fig. \ref{spectra}(h)) is more logically assigned to one of the inner trilobite-like wells with $R\leq40\,a_0$ (Fig. \ref{spectra}(b)).

Several interesting questions remain: why have most of the observations to date been well explained by transitions to $^3\Sigma^+$ trilobite-like states, but not to $^3\Pi$ states, which are also allowed by E1 selection rules? Could some of the unassigned large features in Fig. \ref{spectra} be due to these $^3\Pi$ states? In addition, by what mechanism do these trilobite-like states autoionize, given that Rb$_2^+$ must be produced at short range? Trilobite-like states of Rb$_2$ were previously found to decay both into Rb$^+$ and Rb$_2^+$ ions \cite{bendkowsky09,bendkowsky10,butscher10,butscher11}, while those of Cs$_2$ were found to decay only into Cs$^+$ \cite{tallant12}. The decay mechanism of trilobite-like states is currently not fully understood.

One of the advantages of our bound-bound excitation method is that it allows control over the Franck-Condon overlap of the initial and final states. By choosing the initial vibrational level, an experiment can be tailored to excite predominantly a single well in a trilobite-like PEC, simplifying the spectrum and facilitating its assignment. Furthermore, by varying the outer turning point of the initial state, one can study variations of parameters across wells of the excited-state PEC, such as changes in EDM between wells, or mixing between covalent bonding at short range with trilobite-like bonding at long range.

Our present apparatus is limited in this regard because it cannot build up population in the very highest vibrational levels of the $a\,^3\Sigma_u^+$ or $X\,^1\Sigma_g^+$ state, due to inadvertent photodissociation caused by the photoassociation laser \cite{Huang06}. The $\left | a\,^3\Sigma_u^+, v= 35\right>$ level used for this work, with its outer turning point of 35$\,a_0$, is the highest that we can populate with reasonable numbers. This limitation could be overcome by using, for example, stimulated Raman transfer or magnetoassociation via a Feshbach resonance \cite{kohler06} to populate the uppermost vibrational level. For instance, the last bound levels of $^{85}$Rb$_2$ and $^{87}$Rb$_2$ molecules in the $a\,^3\Sigma_u^+$ state have outer turning points of 73$\,a_0$ and 104$\,a_0$, respectively \footnote{We calculate outer turning points using the $a\,^3\Sigma_u^+$ PEC of Ref. \cite{strauss10} and the \emph{LEVEL8.0} \cite{level8} computer program to solve for the binding energies of levels, and hence their turning points.}. Magnetoassociated molecules, with their large outer turning points and rovibrational-state purity, are expected to be particularly good starting points for the population of trilobite-like states.

In conclusion, we demonstrate the production of low-$n$ trilobite-like states through bound-to-bound excitation of weakly-bound molecules. This approach is relatively simple and should be applicable to other diatomic molecules, including heteronuclear systems, and possibly to the production of trilobite-like trimers \cite{bendkowsky10}. Weakly-bound molecules, with their large outer turning points, can have a significant wavefunction overlap with long-range trilobite-like states. We find that our spectra are consistent with calculated PECs and with previously observed spectra in heat-pipe ovens. The ability to produce trilobite-like states in a low-density collision-free regime, will shed light on the decay mechanism of these fragile molecules.


\begin{acknowledgments}
We thank C. H. Greene for valuable discussions, and the authors of Ref. \cite{greene06} and V. Horvatic for making data available to us. We gratefully acknowledge funding from NSF and AFOSR (MURI).
\end{acknowledgments}

\bibliography{C:/Users/Public/bibliography/bibtex_masterfile_donotdelete}

\begin{thebibliography}{33}%
\makeatletter
\providecommand \@ifxundefined [1]{%
 \@ifx{#1\undefined}
}%
\providecommand \@ifnum [1]{%
 \ifnum #1\expandafter \@firstoftwo
 \else \expandafter \@secondoftwo
 \fi
}%
\providecommand \@ifx [1]{%
 \ifx #1\expandafter \@firstoftwo
 \else \expandafter \@secondoftwo
 \fi
}%
\providecommand \natexlab [1]{#1}%
\providecommand \enquote  [1]{``#1''}%
\providecommand \bibnamefont  [1]{#1}%
\providecommand \bibfnamefont [1]{#1}%
\providecommand \citenamefont [1]{#1}%
\providecommand \href@noop [0]{\@secondoftwo}%
\providecommand \href [0]{\begingroup \@sanitize@url \@href}%
\providecommand \@href[1]{\@@startlink{#1}\@@href}%
\providecommand \@@href[1]{\endgroup#1\@@endlink}%
\providecommand \@sanitize@url [0]{\catcode `\\12\catcode `\$12\catcode
  `\&12\catcode `\#12\catcode `\^12\catcode `\_12\catcode `\%12\relax}%
\providecommand \@@startlink[1]{}%
\providecommand \@@endlink[0]{}%
\providecommand \url  [0]{\begingroup\@sanitize@url \@url }%
\providecommand \@url [1]{\endgroup\@href {#1}{\urlprefix }}%
\providecommand \urlprefix  [0]{URL }%
\providecommand \Eprint [0]{\href }%
\providecommand \doibase [0]{http://dx.doi.org/}%
\providecommand \selectlanguage [0]{\@gobble}%
\providecommand \bibinfo  [0]{\@secondoftwo}%
\providecommand \bibfield  [0]{\@secondoftwo}%
\providecommand \translation [1]{[#1]}%
\providecommand \BibitemOpen [0]{}%
\providecommand \bibitemStop [0]{}%
\providecommand \bibitemNoStop [0]{.\EOS\space}%
\providecommand \EOS [0]{\spacefactor3000\relax}%
\providecommand \BibitemShut  [1]{\csname bibitem#1\endcsname}%
\let\auto@bib@innerbib\@empty
%</preamble>
\bibitem [{\citenamefont {Greene}\ \emph {et~al.}(2000)\citenamefont {Greene},
  \citenamefont {Dickinson},\ and\ \citenamefont {Sadeghpour}}]{greene00}%
  \BibitemOpen
  \bibfield  {author} {\bibinfo {author} {\bibfnamefont {C.~H.}\ \bibnamefont
  {Greene}}, \bibinfo {author} {\bibfnamefont {A.~S.}\ \bibnamefont
  {Dickinson}}, \ and\ \bibinfo {author} {\bibfnamefont {H.~R.}\ \bibnamefont
  {Sadeghpour}},\ }\href {\doibase 10.1103/PhysRevLett.85.2458} {\bibfield
  {journal} {\bibinfo  {journal} {Phys. Rev. Lett.}\ }\textbf {\bibinfo
  {volume} {85}},\ \bibinfo {pages} {2458} (\bibinfo {year}
  {2000})}\BibitemShut {NoStop}%
\bibitem [{\citenamefont {Hogan}\ and\ \citenamefont {Merkt}(2009)}]{hogan09}%
  \BibitemOpen
  \bibfield  {author} {\bibinfo {author} {\bibfnamefont {S.~D.}\ \bibnamefont
  {Hogan}}\ and\ \bibinfo {author} {\bibfnamefont {F.}~\bibnamefont {Merkt}},\
  }\href {\doibase 10.1002/cphc.200900499} {\bibfield  {journal} {\bibinfo
  {journal} {Chem. Phys. Chem.}\ }\textbf {\bibinfo {volume} {10}},\ \bibinfo
  {pages} {2931} (\bibinfo {year} {2009})}\BibitemShut {NoStop}%
\bibitem [{\citenamefont {Bendkowsky}\ \emph {et~al.}(2009)\citenamefont
  {Bendkowsky}, \citenamefont {Butscher}, \citenamefont {Nipper}, \citenamefont
  {Shaffer}, \citenamefont {L\"{o}w},\ and\ \citenamefont
  {Pfau}}]{bendkowsky09}%
  \BibitemOpen
  \bibfield  {author} {\bibinfo {author} {\bibfnamefont {V.}~\bibnamefont
  {Bendkowsky}}, \bibinfo {author} {\bibfnamefont {B.}~\bibnamefont
  {Butscher}}, \bibinfo {author} {\bibfnamefont {J.}~\bibnamefont {Nipper}},
  \bibinfo {author} {\bibfnamefont {J.~P.}\ \bibnamefont {Shaffer}}, \bibinfo
  {author} {\bibfnamefont {R.}~\bibnamefont {L\"{o}w}}, \ and\ \bibinfo
  {author} {\bibfnamefont {T.}~\bibnamefont {Pfau}},\ }\href {\doibase
  dx.doi.org/10.1038/nature07945} {\bibfield  {journal} {\bibinfo  {journal}
  {Nature}\ }\textbf {\bibinfo {volume} {458}},\ \bibinfo {pages} {1005}
  (\bibinfo {year} {2009})}\BibitemShut {NoStop}%
\bibitem [{\citenamefont {Bendkowsky}\ \emph {et~al.}(2010)\citenamefont
  {Bendkowsky}, \citenamefont {Butscher}, \citenamefont {Nipper}, \citenamefont
  {Balewski}, \citenamefont {Shaffer}, \citenamefont {L\"{o}w}, \citenamefont
  {Pfau}, \citenamefont {Li}, \citenamefont {Stanojevic}, \citenamefont
  {Pohl},\ and\ \citenamefont {Rost}}]{bendkowsky10}%
  \BibitemOpen
  \bibfield  {author} {\bibinfo {author} {\bibfnamefont {V.}~\bibnamefont
  {Bendkowsky}}, \bibinfo {author} {\bibfnamefont {B.}~\bibnamefont
  {Butscher}}, \bibinfo {author} {\bibfnamefont {J.}~\bibnamefont {Nipper}},
  \bibinfo {author} {\bibfnamefont {J.~B.}\ \bibnamefont {Balewski}}, \bibinfo
  {author} {\bibfnamefont {J.~P.}\ \bibnamefont {Shaffer}}, \bibinfo {author}
  {\bibfnamefont {R.}~\bibnamefont {L\"{o}w}}, \bibinfo {author} {\bibfnamefont
  {T.}~\bibnamefont {Pfau}}, \bibinfo {author} {\bibfnamefont {W.}~\bibnamefont
  {Li}}, \bibinfo {author} {\bibfnamefont {J.}~\bibnamefont {Stanojevic}},
  \bibinfo {author} {\bibfnamefont {T.}~\bibnamefont {Pohl}}, \ and\ \bibinfo
  {author} {\bibfnamefont {J.~M.}\ \bibnamefont {Rost}},\ }\href {\doibase
  10.1103/PhysRevLett.105.163201} {\bibfield  {journal} {\bibinfo  {journal}
  {Phys. Rev. Lett.}\ }\textbf {\bibinfo {volume} {105}},\ \bibinfo {pages}
  {163201} (\bibinfo {year} {2010})}\BibitemShut {NoStop}%
\bibitem [{\citenamefont {Butscher}\ \emph {et~al.}(2010)\citenamefont
  {Butscher}, \citenamefont {Nipper}, \citenamefont {Balewski}, \citenamefont
  {Kukota}, \citenamefont {Bendkowsky}, \citenamefont {L\"{o}w},\ and\
  \citenamefont {Pfau}}]{butscher10}%
  \BibitemOpen
  \bibfield  {author} {\bibinfo {author} {\bibfnamefont {B.}~\bibnamefont
  {Butscher}}, \bibinfo {author} {\bibfnamefont {J.}~\bibnamefont {Nipper}},
  \bibinfo {author} {\bibfnamefont {J.~B.}\ \bibnamefont {Balewski}}, \bibinfo
  {author} {\bibfnamefont {L.}~\bibnamefont {Kukota}}, \bibinfo {author}
  {\bibfnamefont {V.}~\bibnamefont {Bendkowsky}}, \bibinfo {author}
  {\bibfnamefont {R.}~\bibnamefont {L\"{o}w}}, \ and\ \bibinfo {author}
  {\bibfnamefont {T.}~\bibnamefont {Pfau}},\ }\href {\doibase
  10.1038/nphys1828} {\bibfield  {journal} {\bibinfo  {journal} {Nature Phys.}\
  }\textbf {\bibinfo {volume} {6}},\ \bibinfo {pages} {970} (\bibinfo {year}
  {2010})}\BibitemShut {NoStop}%
\bibitem [{\citenamefont {Butscher}\ \emph {et~al.}(2011)\citenamefont
  {Butscher}, \citenamefont {Bendkowsky}, \citenamefont {Nipper}, \citenamefont
  {Balewski}, \citenamefont {Kukota}, \citenamefont {L\"{o}w}, \citenamefont
  {Pfau}, \citenamefont {Li}, \citenamefont {Pohl},\ and\ \citenamefont
  {Rost}}]{butscher11}%
  \BibitemOpen
  \bibfield  {author} {\bibinfo {author} {\bibfnamefont {B.}~\bibnamefont
  {Butscher}}, \bibinfo {author} {\bibfnamefont {V.}~\bibnamefont
  {Bendkowsky}}, \bibinfo {author} {\bibfnamefont {J.}~\bibnamefont {Nipper}},
  \bibinfo {author} {\bibfnamefont {J.~B.}\ \bibnamefont {Balewski}}, \bibinfo
  {author} {\bibfnamefont {L.}~\bibnamefont {Kukota}}, \bibinfo {author}
  {\bibfnamefont {R.}~\bibnamefont {L\"{o}w}}, \bibinfo {author} {\bibfnamefont
  {T.}~\bibnamefont {Pfau}}, \bibinfo {author} {\bibfnamefont {W.}~\bibnamefont
  {Li}}, \bibinfo {author} {\bibfnamefont {T.}~\bibnamefont {Pohl}}, \ and\
  \bibinfo {author} {\bibfnamefont {J.~M.}\ \bibnamefont {Rost}},\ }\href
  {http://stacks.iop.org/0953-4075/44/i=18/a=184004} {\bibfield  {journal}
  {\bibinfo  {journal} {J. Phys. B}\ }\textbf {\bibinfo {volume} {44}},\
  \bibinfo {pages} {184004} (\bibinfo {year} {2011})}\BibitemShut {NoStop}%
\bibitem [{\citenamefont {Li}\ \emph {et~al.}(2011)\citenamefont {Li},
  \citenamefont {Pohl}, \citenamefont {Rost}, \citenamefont {Rittenhouse},
  \citenamefont {Sadeghpour}, \citenamefont {Nipper}, \citenamefont {Butscher},
  \citenamefont {Balewski}, \citenamefont {Bendkowsky}, \citenamefont
  {L\"{o}w},\ and\ \citenamefont {Pfau}}]{li11}%
  \BibitemOpen
  \bibfield  {author} {\bibinfo {author} {\bibfnamefont {W.}~\bibnamefont
  {Li}}, \bibinfo {author} {\bibfnamefont {T.}~\bibnamefont {Pohl}}, \bibinfo
  {author} {\bibfnamefont {J.~M.}\ \bibnamefont {Rost}}, \bibinfo {author}
  {\bibfnamefont {S.~T.}\ \bibnamefont {Rittenhouse}}, \bibinfo {author}
  {\bibfnamefont {H.~R.}\ \bibnamefont {Sadeghpour}}, \bibinfo {author}
  {\bibfnamefont {J.}~\bibnamefont {Nipper}}, \bibinfo {author} {\bibfnamefont
  {B.}~\bibnamefont {Butscher}}, \bibinfo {author} {\bibfnamefont {J.~B.}\
  \bibnamefont {Balewski}}, \bibinfo {author} {\bibfnamefont {V.}~\bibnamefont
  {Bendkowsky}}, \bibinfo {author} {\bibfnamefont {R.}~\bibnamefont {L\"{o}w}},
  \ and\ \bibinfo {author} {\bibfnamefont {T.}~\bibnamefont {Pfau}},\ }\href
  {\doibase 10.1126/science.1211255} {\bibfield  {journal} {\bibinfo  {journal}
  {Science}\ }\textbf {\bibinfo {volume} {334}},\ \bibinfo {pages} {1110}
  (\bibinfo {year} {2011})}\BibitemShut {NoStop}%
\bibitem [{\citenamefont {Tallant}\ \emph {et~al.}(2012)\citenamefont
  {Tallant}, \citenamefont {Rittenhouse}, \citenamefont {Booth}, \citenamefont
  {Sadeghpour},\ and\ \citenamefont {Shaffer}}]{tallant12}%
  \BibitemOpen
  \bibfield  {author} {\bibinfo {author} {\bibfnamefont {J.}~\bibnamefont
  {Tallant}}, \bibinfo {author} {\bibfnamefont {S.~T.}\ \bibnamefont
  {Rittenhouse}}, \bibinfo {author} {\bibfnamefont {D.}~\bibnamefont {Booth}},
  \bibinfo {author} {\bibfnamefont {H.~R.}\ \bibnamefont {Sadeghpour}}, \ and\
  \bibinfo {author} {\bibfnamefont {J.~P.}\ \bibnamefont {Shaffer}},\ }\href
  {\doibase 10.1103/PhysRevLett.109.173202} {\bibfield  {journal} {\bibinfo
  {journal} {Phys. Rev. Lett.}\ }\textbf {\bibinfo {volume} {109}},\ \bibinfo
  {pages} {173202} (\bibinfo {year} {2012})}\BibitemShut {NoStop}%
\bibitem [{\citenamefont {Greene}\ \emph {et~al.}(2006)\citenamefont {Greene},
  \citenamefont {Hamilton}, \citenamefont {Crowell}, \citenamefont {Vadla},\
  and\ \citenamefont {Niemax}}]{greene06}%
  \BibitemOpen
  \bibfield  {author} {\bibinfo {author} {\bibfnamefont {C.~H.}\ \bibnamefont
  {Greene}}, \bibinfo {author} {\bibfnamefont {E.~L.}\ \bibnamefont
  {Hamilton}}, \bibinfo {author} {\bibfnamefont {H.}~\bibnamefont {Crowell}},
  \bibinfo {author} {\bibfnamefont {C.}~\bibnamefont {Vadla}}, \ and\ \bibinfo
  {author} {\bibfnamefont {K.}~\bibnamefont {Niemax}},\ }\href {\doibase
  10.1103/PhysRevLett.97.233002} {\bibfield  {journal} {\bibinfo  {journal}
  {Phys. Rev. Lett.}\ }\textbf {\bibinfo {volume} {97}},\ \bibinfo {pages}
  {233002} (\bibinfo {year} {2006})}\BibitemShut {NoStop}%
\bibitem [{\citenamefont {Vadla}\ \emph {et~al.}(2009)\citenamefont {Vadla},
  \citenamefont {Horvatic},\ and\ \citenamefont {Niemax}}]{vadla09}%
  \BibitemOpen
  \bibfield  {author} {\bibinfo {author} {\bibfnamefont {C.}~\bibnamefont
  {Vadla}}, \bibinfo {author} {\bibfnamefont {V.}~\bibnamefont {Horvatic}}, \
  and\ \bibinfo {author} {\bibfnamefont {K.}~\bibnamefont {Niemax}},\ }\href
  {\doibase 10.1103/PhysRevA.80.052506} {\bibfield  {journal} {\bibinfo
  {journal} {Phys. Rev. A}\ }\textbf {\bibinfo {volume} {80}},\ \bibinfo
  {pages} {052506} (\bibinfo {year} {2009})}\BibitemShut {NoStop}%
\bibitem [{\citenamefont {Boisseau}\ \emph {et~al.}(2002)\citenamefont
  {Boisseau}, \citenamefont {Simbotin},\ and\ \citenamefont
  {C\^ot\'e}}]{boisseau02}%
  \BibitemOpen
  \bibfield  {author} {\bibinfo {author} {\bibfnamefont {C.}~\bibnamefont
  {Boisseau}}, \bibinfo {author} {\bibfnamefont {I.}~\bibnamefont {Simbotin}},
  \ and\ \bibinfo {author} {\bibfnamefont {R.}~\bibnamefont {C\^ot\'e}},\
  }\href {\doibase 10.1103/PhysRevLett.88.133004} {\bibfield  {journal}
  {\bibinfo  {journal} {Phys. Rev. Lett.}\ }\textbf {\bibinfo {volume} {88}},\
  \bibinfo {pages} {133004} (\bibinfo {year} {2002})}\BibitemShut {NoStop}%
\bibitem [{\citenamefont {Farooqi}\ \emph {et~al.}(2003)\citenamefont
  {Farooqi}, \citenamefont {Tong}, \citenamefont {Krishnan}, \citenamefont
  {Stanojevic}, \citenamefont {Zhang}, \citenamefont {Ensher}, \citenamefont
  {Estrin}, \citenamefont {Boisseau}, \citenamefont {C\^ot\'e}, \citenamefont
  {Eyler},\ and\ \citenamefont {Gould}}]{farooqi03}%
  \BibitemOpen
  \bibfield  {author} {\bibinfo {author} {\bibfnamefont {S.~M.}\ \bibnamefont
  {Farooqi}}, \bibinfo {author} {\bibfnamefont {D.}~\bibnamefont {Tong}},
  \bibinfo {author} {\bibfnamefont {S.}~\bibnamefont {Krishnan}}, \bibinfo
  {author} {\bibfnamefont {J.}~\bibnamefont {Stanojevic}}, \bibinfo {author}
  {\bibfnamefont {Y.~P.}\ \bibnamefont {Zhang}}, \bibinfo {author}
  {\bibfnamefont {J.~R.}\ \bibnamefont {Ensher}}, \bibinfo {author}
  {\bibfnamefont {A.~S.}\ \bibnamefont {Estrin}}, \bibinfo {author}
  {\bibfnamefont {C.}~\bibnamefont {Boisseau}}, \bibinfo {author}
  {\bibfnamefont {R.}~\bibnamefont {C\^ot\'e}}, \bibinfo {author}
  {\bibfnamefont {E.~E.}\ \bibnamefont {Eyler}}, \ and\ \bibinfo {author}
  {\bibfnamefont {P.~L.}\ \bibnamefont {Gould}},\ }\href {\doibase
  10.1103/PhysRevLett.91.183002} {\bibfield  {journal} {\bibinfo  {journal}
  {Phys. Rev. Lett.}\ }\textbf {\bibinfo {volume} {91}},\ \bibinfo {pages}
  {183002} (\bibinfo {year} {2003})}\BibitemShut {NoStop}%
\bibitem [{\citenamefont {Overstreet}\ \emph {et~al.}(2009)\citenamefont
  {Overstreet}, \citenamefont {Schwettmann}, \citenamefont {Tallant},
  \citenamefont {Booth},\ and\ \citenamefont {Shaffer}}]{overstreet09}%
  \BibitemOpen
  \bibfield  {author} {\bibinfo {author} {\bibfnamefont {K.~R.}\ \bibnamefont
  {Overstreet}}, \bibinfo {author} {\bibfnamefont {A.}~\bibnamefont
  {Schwettmann}}, \bibinfo {author} {\bibfnamefont {J.}~\bibnamefont
  {Tallant}}, \bibinfo {author} {\bibfnamefont {D.}~\bibnamefont {Booth}}, \
  and\ \bibinfo {author} {\bibfnamefont {J.~P.}\ \bibnamefont {Shaffer}},\
  }\href {\doibase 10.1038/nphys1307} {\bibfield  {journal} {\bibinfo
  {journal} {Nature Phys.}\ }\textbf {\bibinfo {volume} {5}},\ \bibinfo {pages}
  {581} (\bibinfo {year} {2009})}\BibitemShut {NoStop}%
\bibitem [{\citenamefont {Reinhold}\ and\ \citenamefont
  {Ubachs}(2005)}]{reinhold05}%
  \BibitemOpen
  \bibfield  {author} {\bibinfo {author} {\bibfnamefont {E.}~\bibnamefont
  {Reinhold}}\ and\ \bibinfo {author} {\bibfnamefont {W.}~\bibnamefont
  {Ubachs}},\ }\href {\doibase 10.1080/00268970500050621} {\bibfield  {journal}
  {\bibinfo  {journal} {Mol. Phys.}\ }\textbf {\bibinfo {volume} {103}},\
  \bibinfo {pages} {1329} (\bibinfo {year} {2005})}\BibitemShut {NoStop}%
\bibitem [{\citenamefont {Kirrander}\ \emph {et~al.}(2013)\citenamefont
  {Kirrander}, \citenamefont {Rittenhouse}, \citenamefont {Ascoli},
  \citenamefont {Eyler}, \citenamefont {Gould},\ and\ \citenamefont
  {Sadeghpour}}]{kirrander13}%
  \BibitemOpen
  \bibfield  {author} {\bibinfo {author} {\bibfnamefont {A.}~\bibnamefont
  {Kirrander}}, \bibinfo {author} {\bibfnamefont {S.}~\bibnamefont
  {Rittenhouse}}, \bibinfo {author} {\bibfnamefont {M.}~\bibnamefont {Ascoli}},
  \bibinfo {author} {\bibfnamefont {E.~E.}\ \bibnamefont {Eyler}}, \bibinfo
  {author} {\bibfnamefont {P.~L.}\ \bibnamefont {Gould}}, \ and\ \bibinfo
  {author} {\bibfnamefont {H.~R.}\ \bibnamefont {Sadeghpour}},\ }\href
  {\doibase 10.1103/PhysRevA.87.031402} {\bibfield  {journal} {\bibinfo
  {journal} {Phys. Rev. A}\ }\textbf {\bibinfo {volume} {87}},\ \bibinfo
  {pages} {031402} (\bibinfo {year} {2013})}\BibitemShut {NoStop}%
\bibitem [{\citenamefont {Omont}(1977)}]{omont77}%
  \BibitemOpen
  \bibfield  {author} {\bibinfo {author} {\bibfnamefont {A.}~\bibnamefont
  {Omont}},\ }\href {\doibase 10.1051/jphys:0197700380110134300} {\bibfield
  {journal} {\bibinfo  {journal} {J. Phys. France}\ }\textbf {\bibinfo {volume}
  {38}},\ \bibinfo {pages} {1343} (\bibinfo {year} {1977})}\BibitemShut
  {NoStop}%
\bibitem [{\citenamefont {Hamilton}\ \emph {et~al.}(2002)\citenamefont
  {Hamilton}, \citenamefont {Greene},\ and\ \citenamefont
  {Sadeghpour}}]{hamilton02}%
  \BibitemOpen
  \bibfield  {author} {\bibinfo {author} {\bibfnamefont {E.~L.}\ \bibnamefont
  {Hamilton}}, \bibinfo {author} {\bibfnamefont {C.~H.}\ \bibnamefont
  {Greene}}, \ and\ \bibinfo {author} {\bibfnamefont {H.~R.}\ \bibnamefont
  {Sadeghpour}},\ }\href {http://stacks.iop.org/0953-4075/35/i=10/a=102}
  {\bibfield  {journal} {\bibinfo  {journal} {J. Phys. B}\ }\textbf {\bibinfo
  {volume} {35}},\ \bibinfo {pages} {L199} (\bibinfo {year}
  {2002})}\BibitemShut {NoStop}%
\bibitem [{\citenamefont {Bahrim}\ and\ \citenamefont
  {Thumm}(2000)}]{bahrim00}%
  \BibitemOpen
  \bibfield  {author} {\bibinfo {author} {\bibfnamefont {C.}~\bibnamefont
  {Bahrim}}\ and\ \bibinfo {author} {\bibfnamefont {U.}~\bibnamefont {Thumm}},\
  }\href {\doibase 10.1103/PhysRevA.61.022722} {\bibfield  {journal} {\bibinfo
  {journal} {Phys. Rev. A}\ }\textbf {\bibinfo {volume} {61}},\ \bibinfo
  {pages} {022722} (\bibinfo {year} {2000})}\BibitemShut {NoStop}%
\bibitem [{\citenamefont {Khuskivadze}\ \emph {et~al.}(2002)\citenamefont
  {Khuskivadze}, \citenamefont {Chibisov},\ and\ \citenamefont
  {Fabrikant}}]{khuskivadze02}%
  \BibitemOpen
  \bibfield  {author} {\bibinfo {author} {\bibfnamefont {A.~A.}\ \bibnamefont
  {Khuskivadze}}, \bibinfo {author} {\bibfnamefont {M.~I.}\ \bibnamefont
  {Chibisov}}, \ and\ \bibinfo {author} {\bibfnamefont {I.~I.}\ \bibnamefont
  {Fabrikant}},\ }\href {\doibase 10.1103/PhysRevA.66.042709} {\bibfield
  {journal} {\bibinfo  {journal} {Phys. Rev. A}\ }\textbf {\bibinfo {volume}
  {66}},\ \bibinfo {pages} {042709} (\bibinfo {year} {2002})}\BibitemShut
  {NoStop}%
\bibitem [{\citenamefont {Junginger}\ \emph {et~al.}(2012)\citenamefont
  {Junginger}, \citenamefont {Main},\ and\ \citenamefont
  {Wunner}}]{junginger12}%
  \BibitemOpen
  \bibfield  {author} {\bibinfo {author} {\bibfnamefont {A.}~\bibnamefont
  {Junginger}}, \bibinfo {author} {\bibfnamefont {J.}~\bibnamefont {Main}}, \
  and\ \bibinfo {author} {\bibfnamefont {G.}~\bibnamefont {Wunner}},\ }\href
  {\doibase 10.1103/PhysRevA.86.012713} {\bibfield  {journal} {\bibinfo
  {journal} {Phys. Rev. A}\ }\textbf {\bibinfo {volume} {86}},\ \bibinfo
  {pages} {012713} (\bibinfo {year} {2012})}\BibitemShut {NoStop}%
\bibitem [{\citenamefont {Yiannopoulou}\ \emph {et~al.}(1999)\citenamefont
  {Yiannopoulou}, \citenamefont {Jeung}, \citenamefont {Park}, \citenamefont
  {Lee},\ and\ \citenamefont {Lee}}]{yiannopoulou99}%
  \BibitemOpen
  \bibfield  {author} {\bibinfo {author} {\bibfnamefont {A.}~\bibnamefont
  {Yiannopoulou}}, \bibinfo {author} {\bibfnamefont {G.-H.}\ \bibnamefont
  {Jeung}}, \bibinfo {author} {\bibfnamefont {S.~J.}\ \bibnamefont {Park}},
  \bibinfo {author} {\bibfnamefont {H.~S.}\ \bibnamefont {Lee}}, \ and\
  \bibinfo {author} {\bibfnamefont {Y.~S.}\ \bibnamefont {Lee}},\ }\href
  {\doibase 10.1103/PhysRevA.59.1178} {\bibfield  {journal} {\bibinfo
  {journal} {Phys. Rev. A}\ }\textbf {\bibinfo {volume} {59}},\ \bibinfo
  {pages} {1178} (\bibinfo {year} {1999})}\BibitemShut {NoStop}%
\bibitem [{Note1()}]{Note1}%
  \BibitemOpen
  \bibinfo {note} {For the ground-state potential, $V(R)\protect \tmspace
  +\thinmuskip {.1667em}=\protect \tmspace +\thinmuskip {.1667em}-\protect
  \frac {C_6}{R^6}$, $C_6\protect \tmspace +\thinmuskip {.1667em}=\protect
  \tmspace +\thinmuskip {.1667em}4660$ a.u., see Ref. \cite {greene06}. The
  resulting correction between difference PECs and regular PECs is less than
  0.5$\protect \tmspace +\thinmuskip {.1667em}$cm$^{-1}$ at $R=35\protect
  \tmspace +\thinmuskip {.1667em}a_0$.}\BibitemShut {Stop}%
\bibitem [{\citenamefont {Allouche}\ and\ \citenamefont
  {Aubert-Fr\'{e}con}(2012)}]{allouche12}%
  \BibitemOpen
  \bibfield  {author} {\bibinfo {author} {\bibfnamefont {A.-R.}\ \bibnamefont
  {Allouche}}\ and\ \bibinfo {author} {\bibfnamefont {M.}~\bibnamefont
  {Aubert-Fr\'{e}con}},\ }\href {\doibase 10.1063/1.3694014} {\bibfield
  {journal} {\bibinfo  {journal} {J. Chem. Phys.}\ }\textbf {\bibinfo {volume}
  {136}},\ \bibinfo {eid} {114302} (\bibinfo {year} {2012})}\BibitemShut
  {NoStop}%
\bibitem [{\citenamefont {Bellos}\ \emph {et~al.}(2013)\citenamefont {Bellos},
  \citenamefont {Carollo}, \citenamefont {Banerjee}, \citenamefont {Ascoli},
  \citenamefont {Allouche}, \citenamefont {Eyler}, \citenamefont {Gould},\ and\
  \citenamefont {Stwalley}}]{bellos13}%
  \BibitemOpen
  \bibfield  {author} {\bibinfo {author} {\bibfnamefont {M.~A.}\ \bibnamefont
  {Bellos}}, \bibinfo {author} {\bibfnamefont {R.}~\bibnamefont {Carollo}},
  \bibinfo {author} {\bibfnamefont {J.}~\bibnamefont {Banerjee}}, \bibinfo
  {author} {\bibfnamefont {M.}~\bibnamefont {Ascoli}}, \bibinfo {author}
  {\bibfnamefont {A.-R.}\ \bibnamefont {Allouche}}, \bibinfo {author}
  {\bibfnamefont {E.~E.}\ \bibnamefont {Eyler}}, \bibinfo {author}
  {\bibfnamefont {P.~L.}\ \bibnamefont {Gould}}, \ and\ \bibinfo {author}
  {\bibfnamefont {W.~C.}\ \bibnamefont {Stwalley}},\ }\href {\doibase
  10.1103/PhysRevA.87.012508} {\bibfield  {journal} {\bibinfo  {journal} {Phys.
  Rev. A}\ }\textbf {\bibinfo {volume} {87}},\ \bibinfo {pages} {012508}
  (\bibinfo {year} {2013})}\BibitemShut {NoStop}%
\bibitem [{\citenamefont {Bhatti}\ \emph {et~al.}(1981)\citenamefont {Bhatti},
  \citenamefont {Cromer},\ and\ \citenamefont {Cooke}}]{bhatti81}%
  \BibitemOpen
  \bibfield  {author} {\bibinfo {author} {\bibfnamefont {S.~A.}\ \bibnamefont
  {Bhatti}}, \bibinfo {author} {\bibfnamefont {C.~L.}\ \bibnamefont {Cromer}},
  \ and\ \bibinfo {author} {\bibfnamefont {W.~E.}\ \bibnamefont {Cooke}},\
  }\href {\doibase 10.1103/PhysRevA.24.161} {\bibfield  {journal} {\bibinfo
  {journal} {Phys. Rev. A}\ }\textbf {\bibinfo {volume} {24}},\ \bibinfo
  {pages} {161} (\bibinfo {year} {1981})}\BibitemShut {NoStop}%
\bibitem [{\citenamefont {Lozeille}\ \emph {et~al.}(2006)\citenamefont
  {Lozeille}, \citenamefont {Fioretti}, \citenamefont {Gabbanini},
  \citenamefont {Huang}, \citenamefont {Pechkis}, \citenamefont {Wang},
  \citenamefont {Gould}, \citenamefont {Eyler}, \citenamefont {Stwalley},
  \citenamefont {Aymar},\ and\ \citenamefont {Dulieu}}]{lozeille06}%
  \BibitemOpen
  \bibfield  {author} {\bibinfo {author} {\bibfnamefont {J.}~\bibnamefont
  {Lozeille}}, \bibinfo {author} {\bibfnamefont {A.}~\bibnamefont {Fioretti}},
  \bibinfo {author} {\bibfnamefont {C.}~\bibnamefont {Gabbanini}}, \bibinfo
  {author} {\bibfnamefont {Y.}~\bibnamefont {Huang}}, \bibinfo {author}
  {\bibfnamefont {H.~K.}\ \bibnamefont {Pechkis}}, \bibinfo {author}
  {\bibfnamefont {D.}~\bibnamefont {Wang}}, \bibinfo {author} {\bibfnamefont
  {P.~L.}\ \bibnamefont {Gould}}, \bibinfo {author} {\bibfnamefont {E.~E.}\
  \bibnamefont {Eyler}}, \bibinfo {author} {\bibfnamefont {W.~C.}\ \bibnamefont
  {Stwalley}}, \bibinfo {author} {\bibfnamefont {M.}~\bibnamefont {Aymar}}, \
  and\ \bibinfo {author} {\bibfnamefont {O.}~\bibnamefont {Dulieu}},\ }\href
  {http://dx.doi.org/10.1140/epjd/e2006-00084-4} {\bibfield  {journal}
  {\bibinfo  {journal} {Eur. Phys. J. D}\ }\textbf {\bibinfo {volume} {39}},\
  \bibinfo {pages} {261} (\bibinfo {year} {2006})}\BibitemShut {NoStop}%
\bibitem [{\citenamefont {Salour}(1977)}]{salour77}%
  \BibitemOpen
  \bibfield  {author} {\bibinfo {author} {\bibfnamefont {M.}~\bibnamefont
  {Salour}},\ }\href {\doibase 10.1016/0030-4018(77)90019-0} {\bibfield
  {journal} {\bibinfo  {journal} {Opt. Commun.}\ }\textbf {\bibinfo {volume}
  {22}},\ \bibinfo {pages} {202 } (\bibinfo {year} {1977})}\BibitemShut
  {NoStop}%
\bibitem [{\citenamefont {Eyler}\ \emph {et~al.}(1997)\citenamefont {Eyler},
  \citenamefont {Yiannopoulou}, \citenamefont {Gangopadhyay},\ and\
  \citenamefont {Melikechi}}]{eyler97}%
  \BibitemOpen
  \bibfield  {author} {\bibinfo {author} {\bibfnamefont {E.~E.}\ \bibnamefont
  {Eyler}}, \bibinfo {author} {\bibfnamefont {A.}~\bibnamefont {Yiannopoulou}},
  \bibinfo {author} {\bibfnamefont {S.}~\bibnamefont {Gangopadhyay}}, \ and\
  \bibinfo {author} {\bibfnamefont {N.}~\bibnamefont {Melikechi}},\ }\href
  {\doibase 10.1364/OL.22.000049} {\bibfield  {journal} {\bibinfo  {journal}
  {Opt. Lett.}\ }\textbf {\bibinfo {volume} {22}},\ \bibinfo {pages} {49}
  (\bibinfo {year} {1997})}\BibitemShut {NoStop}%
\bibitem [{\citenamefont {Huang}\ \emph {et~al.}(2006)\citenamefont {Huang},
  \citenamefont {Qi}, \citenamefont {Pechkis}, \citenamefont {Wang},
  \citenamefont {Eyler}, \citenamefont {Gould},\ and\ \citenamefont
  {Stwalley}}]{Huang06}%
  \BibitemOpen
  \bibfield  {author} {\bibinfo {author} {\bibfnamefont {Y.}~\bibnamefont
  {Huang}}, \bibinfo {author} {\bibfnamefont {J.}~\bibnamefont {Qi}}, \bibinfo
  {author} {\bibfnamefont {H.~K.}\ \bibnamefont {Pechkis}}, \bibinfo {author}
  {\bibfnamefont {D.}~\bibnamefont {Wang}}, \bibinfo {author} {\bibfnamefont
  {E.~E.}\ \bibnamefont {Eyler}}, \bibinfo {author} {\bibfnamefont {P.~L.}\
  \bibnamefont {Gould}}, \ and\ \bibinfo {author} {\bibfnamefont {W.~C.}\
  \bibnamefont {Stwalley}},\ }\href
  {http://stacks.iop.org/0953-4075/39/i=19/a=S04} {\bibfield  {journal}
  {\bibinfo  {journal} {J. Phys. B}\ }\textbf {\bibinfo {volume} {39}},\
  \bibinfo {pages} {S857} (\bibinfo {year} {2006})}\BibitemShut {NoStop}%
\bibitem [{\citenamefont {K\"{o}hler}\ \emph {et~al.}(2006)\citenamefont
  {K\"{o}hler}, \citenamefont {G\'{o}ral},\ and\ \citenamefont
  {Julienne}}]{kohler06}%
  \BibitemOpen
  \bibfield  {author} {\bibinfo {author} {\bibfnamefont {T.}~\bibnamefont
  {K\"{o}hler}}, \bibinfo {author} {\bibfnamefont {K.}~\bibnamefont
  {G\'{o}ral}}, \ and\ \bibinfo {author} {\bibfnamefont {P.~S.}\ \bibnamefont
  {Julienne}},\ }\href {\doibase 10.1103/RevModPhys.78.1311} {\bibfield
  {journal} {\bibinfo  {journal} {Rev. Mod. Phys.}\ }\textbf {\bibinfo {volume}
  {78}},\ \bibinfo {pages} {1311} (\bibinfo {year} {2006})}\BibitemShut
  {NoStop}%
\bibitem [{Note2()}]{Note2}%
  \BibitemOpen
  \bibinfo {note} {We calculate outer turning points using the $a\protect
  \tmspace +\thinmuskip {.1667em}^3\Sigma _u^+$ PEC of Ref. \cite {strauss10}
  and the \protect \emph {LEVEL8.0} \cite {level8} computer program to solve
  for the binding energies of levels, and hence their turning
  points.}\BibitemShut {Stop}%
\bibitem [{\citenamefont {Strauss}\ \emph {et~al.}(2010)\citenamefont
  {Strauss}, \citenamefont {Takekoshi}, \citenamefont {Lang}, \citenamefont
  {Winkler}, \citenamefont {Grimm}, \citenamefont {Hecker~Denschlag},\ and\
  \citenamefont {Tiemann}}]{strauss10}%
  \BibitemOpen
  \bibfield  {author} {\bibinfo {author} {\bibfnamefont {C.}~\bibnamefont
  {Strauss}}, \bibinfo {author} {\bibfnamefont {T.}~\bibnamefont {Takekoshi}},
  \bibinfo {author} {\bibfnamefont {F.}~\bibnamefont {Lang}}, \bibinfo {author}
  {\bibfnamefont {K.}~\bibnamefont {Winkler}}, \bibinfo {author} {\bibfnamefont
  {R.}~\bibnamefont {Grimm}}, \bibinfo {author} {\bibfnamefont
  {J.}~\bibnamefont {Hecker~Denschlag}}, \ and\ \bibinfo {author}
  {\bibfnamefont {E.}~\bibnamefont {Tiemann}},\ }\href {\doibase
  10.1103/PhysRevA.82.052514} {\bibfield  {journal} {\bibinfo  {journal} {Phys.
  Rev. A}\ }\textbf {\bibinfo {volume} {82}},\ \bibinfo {pages} {052514}
  (\bibinfo {year} {2010})}\BibitemShut {NoStop}%
\bibitem [{\citenamefont {Le{$\:$}Roy}(2007)}]{level8}%
  \BibitemOpen
  \bibfield  {author} {\bibinfo {author} {\bibfnamefont {R.~J.}\ \bibnamefont
  {Le{$\:$}Roy}},\ }\href {http://scienide2.uwaterloo.ca/{~}rleroy/level/} {\
  (\bibinfo {year} {2007})},\ \bibinfo {note} {{\emph{LEVEL8.0}}: A Computer
  Program for Solving the Radial Schr{\"{o}}dinger Equation for Bound and
  Quasibound Levels, {U}niversity of {W}aterloo Chemical Physics Research
  Report {CP-663} (see
  http://scienide2.uwaterloo.ca/~rleroy/level/)}\BibitemShut {NoStop}%
\end{thebibliography}%

\end{document}